# Ultra-High-Efficiency Dual-Band Thin-Film Lithium Niobate Modulator Incorporating Low-k Underfill with 220 GHz Extrapolated Bandwidth for 390 Gbit/s PAM8 Transmission

Hao Liu[†], Yutong He[†], Bing Xiong[*], Changzheng Sun, Zhibiao Hao, Lai Wang, Jian Wang, Yanjun Han, Hongtao Li, Lin Gan, Yi Luo

*Department of Electronic Engineering, Tsinghua University, Beijing 100084, China*

*† These authors contributed equally to this paper.*

*\*E-mail address: bxiong@tsinghua.edu.cn*

**Abstract:**

High-performance electro-optic modulators play a critical role in modern telecommunication networks and intra-datacenter interconnects. Low driving voltage, large electro-optic bandwidth, compact device size, and multi-band operation ability are essential for various application scenarios, especially energy-efficient high-speed data transmission. However, it is challenging to meet all these requirements simultaneously. Here, we demonstrate a high-performance dual-band thin-film lithium niobate electro-optic modulator with low-k underfill to achieve overall performance improvement. The low-k material helps reduce the RF loss of the modulator and achieve perfect velocity matching with narrow electrode gap to overcome the voltage-bandwidth limitation, extending electro-optic bandwidth and enhancing modulation efficiency simultaneously. The fabricated 7-mm-long modulator exhibits a low half-wave voltage $V_\pi$ of 1.9 V at C-band and 1.54 V at O-band, featuring a low half-wave voltage–length product of 1.33 V·cm and 1.08 V·cm, respectively. Meanwhile, the novel design yields an ultra-wide extrapolated 3 dB bandwidth of 220 GHz (218 GHz) in the C-band (O-band). High-speed data transmission in both C- and O-bands using the same device has been demonstrated for the first time by PAM8 with data rates up to 390 Gbit/s, corresponding to a record-low energy consumption of 0.69 fJ/bit for next-generation cost-effective ultra-high-speed optical communications.

## 1. Introduction

Thin-film lithium niobate (TFLN) has emerged as a versatile photonic platform thanks to its outstanding material properties, including significant Pockels effect (~30.9 pm/V), excellent temperature stability, a wide transparency window (350 nm – 5 μm) and tight optical confinement for compact electrode layout [1]. As the crucial components of photonic integrated circuits, high-speed TFLN electro-optic (EO) modulators with low driving voltage, wide bandwidth and compact size are essential for future high-capacity data centers, telecommunication systems, and quantum optics [2].

Since their advent in 2018 [3], impressive progress has been made in TFLN modulators over the past few years. TFLN modulators based on traditional coplanar waveguide (CPW) electrodes [4–7] are difficult to secure both low half-wave voltage and wide modulation bandwidth simultaneously, as reduced electrode gap leads to increased microwave loss. To overcome the voltage-bandwidth limit of traditional CPW electrodes, Liu et al. proposed employing capacitively-loaded traveling-wave electrodes (CL-TWEs) to improve the performance of TFLN modulators [8]. EO bandwidth over 50 GHz with voltage-length product $V_\pi L$ of 2.6 V·cm [9] and bandwidth exceeding 67 GHz with $V_\pi L$ of 1.7 V·cm [10] are subsequently implemented. Recently, dual-polarization TFLN in-phase quadrature modulator employing CL-TWEs has demonstrated bandwidth up to 110 GHz and $V_\pi L$ of 2.35 V·cm [11]. Special electrode and waveguide structures have been proposed to further enhance the modulation efficiency of TFLN modulators, such as dual-capacitor electrode layout [12] or high-permittivity cladding [13]. However, the former suffers from a low bandwidth due to severe microwave loss and significant velocity mismatch, while the glycerol cladding employed in the latter raises reliability concern. Evidently, achieving high modulation efficiency while maintaining wide EO bandwidth remains a challenge.

Here, we propose and demonstrate an ultra-high-efficiency TFLN Mach-Zehnder modulator (MZM) with low-k benzocyclobutene (BCB) underfill to break the voltage-bandwidth limit and secure significant overall performance improvement. The simple low-k underfill allows for new design flexibility and greatly reduces the microwave loss at high-frequencies while maintaining perfect velocity matching. As a result, narrow electrode gap can be employed for enhanced modulation efficiency, without resorting to complex electrode or waveguide structures. The fabricated with a 7-mm-long TFLN modulator exhibits a low half-wave voltage $V_\pi$ of 1.9 V in the C-band (1550 nm) and 1.54 V in the O-band (1310 nm), featuring an ultra-low $V_\pi L$ of 1.33 V·cm and 1.08 V·cm, respectively. The roll-off in EO frequency response is only 0.77 dB (0.83 dB) up to 110 GHz, corresponding to an extrapolated 3 dB bandwidth of 220 GHz (218 GHz) in the C-band (O-band). High-speed data transmission adopting eight-level pulse-amplitude modulation (PAM8) has demonstrated data rates up to 390 Gbit/s at 130 Gbaud in both C- and O-bands with a record-low energy consumption of 0.69 fJ/bit for intensity modulation and direct detection (IM-DD) system. Compared with other TFLN MZMs reported thus far, our device demonstrates notably higher modulation efficiency while maintaining excellent high-frequency performance, making it a promising candidate for next-generation high-capacity and cost-effective intra-datacenter interconnects and optical communication systems.

**2. Design of High-performance Modulator with Low-k Underfill**

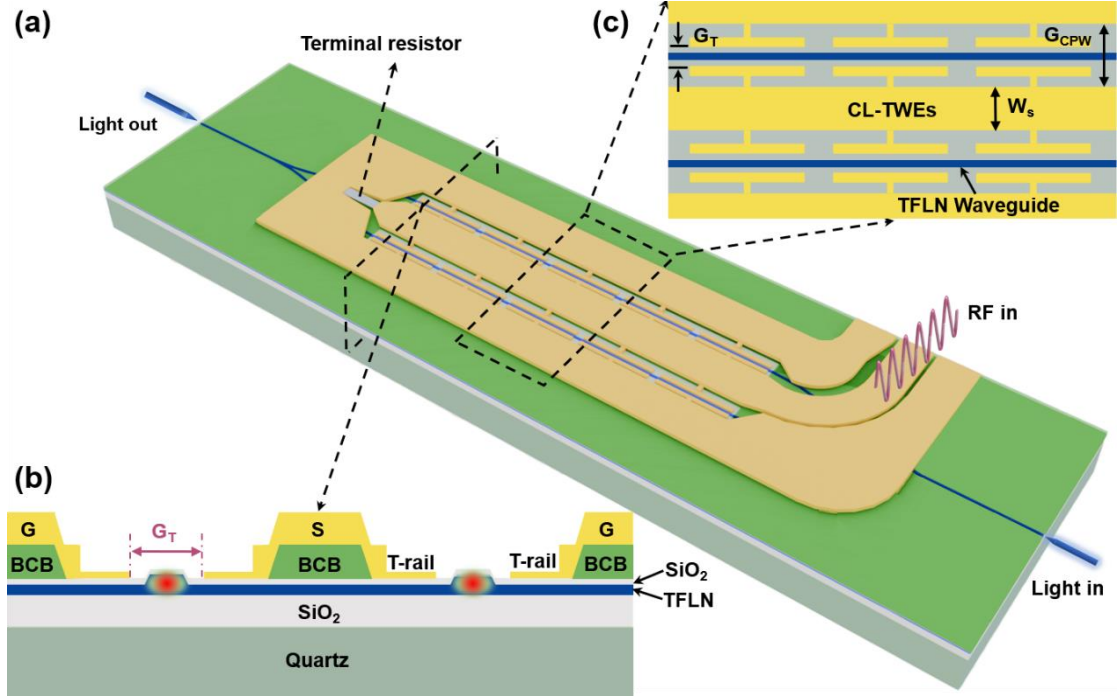

Figure 1. (a) 3D Schematic diagram of the proposed TFLN modulator. (b) The cross-sectional view of the modulation area. (c) Details of the CL-TWEs.

Figure 1 shows the schematic of the proposed TFLN modulator employing CL-TWEs with low-k underfill. The modulator is fabricated on a lithium niobate on insulator (LNOI) wafer, which is composed of a 600-nm-thick X-cut TFLN layer and a 2-μm-thick buried $SiO_2$ layer on a 500-μm-thick quartz substrate. A 100-nm-thick $SiO_2$ buffer layer is deposited beneath the entire electrodes to reduce the optical absorption loss due to the metal electrodes [10]. The etching depth of the LN ridge waveguide is chosen to be 200 nm, while the width and the sidewall angle of the waveguide is taken to be 1 μm and 50°, respectively. The waveguide geometry ensures similar group indices at the C- and O-bands ($n_g$ = 2.221@1550nm and 2.225@1310nm), thus facilitating simultaneous velocity matching at both bands. The RF modulation signal is fed to the modulator via the 90-degree bent GSG pads, and terminated by an on-chip matching resistor at the end of the CL-TWEs.

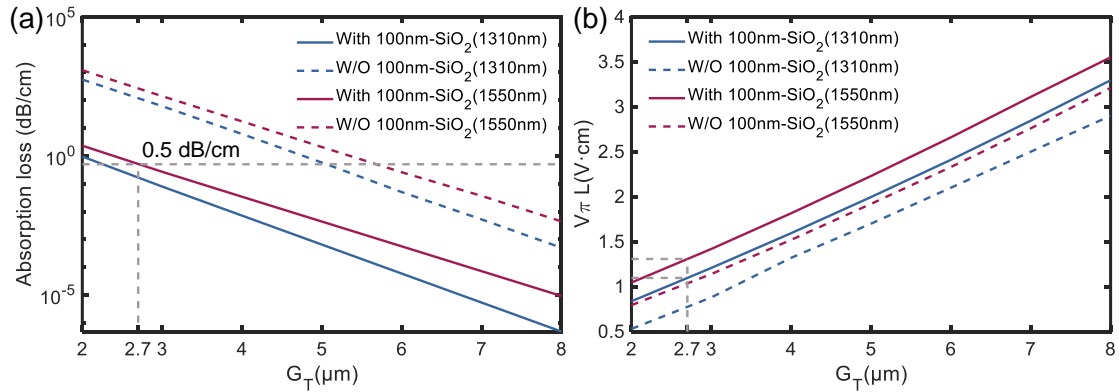

Figure 2. Simulated (a) metal induced optical loss and (b) modulation efficiency with and without 100-nm-thick $SiO_2$ buffer layer in the C- and O-bands.

The performance of the modulator critically depends on the electrode gap between the T-rails of the CL-TWEs. Figure 2 illustrates the influence of electrode gap on metal induced optical absorption loss and modulation efficiency. It is evident that the 100-nm-thick $SiO_2$ buffer layer reduces the optical loss by more than two-orders of magnitude, while increases $V_\pi L$ only slightly. For an electrode gap of 2.7 μm, which ensures an optical absorption loss below 0.5 dB/cm, the modulation efficiency is estimated to be $V_\pi L \sim 1.3$ V·cm (1.09 V·cm) in the C-band (O-band). The higher modulation efficiency at the O band is a result of the enhanced phase accumulation at shorter wavelengths.

The modulation performance of the TFLN modulator at high frequencies is mainly affected by three factors: velocity mismatch between microwave and optical signals, impedance mismatch, and RF loss of the electrodes [14]. RF loss will ultimately limit the modulation bandwidth when velocity and impedance matching can be ensured. RF loss of the modulator mainly comes from the conductor loss of the traveling-wave electrodes, the dielectric loss of the insulating material, and the radiation loss. The conductor loss of the CPW electrodes is given by [15]

$$\alpha_c = k \frac{R_s(f)}{Z_0} = kc_0 R_s(f) \sqrt{\varepsilon_{eff}} C^a \qquad (3)$$

where $c_0$ is the light speed in vacuum, k is an electrode geometry dependent parameter, $R_s(f)$ denotes the surface resistance of the conductors, $\varepsilon_{eff}$ is the effective dielectric constant of the CPW and $C^a$ denotes the electrode structure dependent capacitance in vacuum.

For a given T-rail electrode gap ($G_T$), the width of the signal electrode ($W_S$) and the gap between the main CPW electrodes ($G_{CPW}$) can be adjusted to minimize the velocity and impedance mismatch. The resultant microwave refractive indices and impedances of CL-TWEs with different T-rail gaps are plotted in Figures 3(b) and 3(c), respectively. As revealed by Figure 3(a), reducing the T-rail gap improves the modulation efficiency, but leads to increased microwave loss. This can be explained with the help of Figure 3(d). A narrow T-rail gap corresponds to increased electrode capacitance, which indicates enhanced microwave loss according to Equation (3). Meanwhile, a narrower signal electrode is required to accommodate the increased capacitance for impedance matching, which also contributes to increased microwave loss. Furthermore, according to Figure 3(b), the microwave refractive index increases with reduced T-rail gap, which impedes velocity matching. This poses a compromise between modulation efficiency and modulation bandwidth.

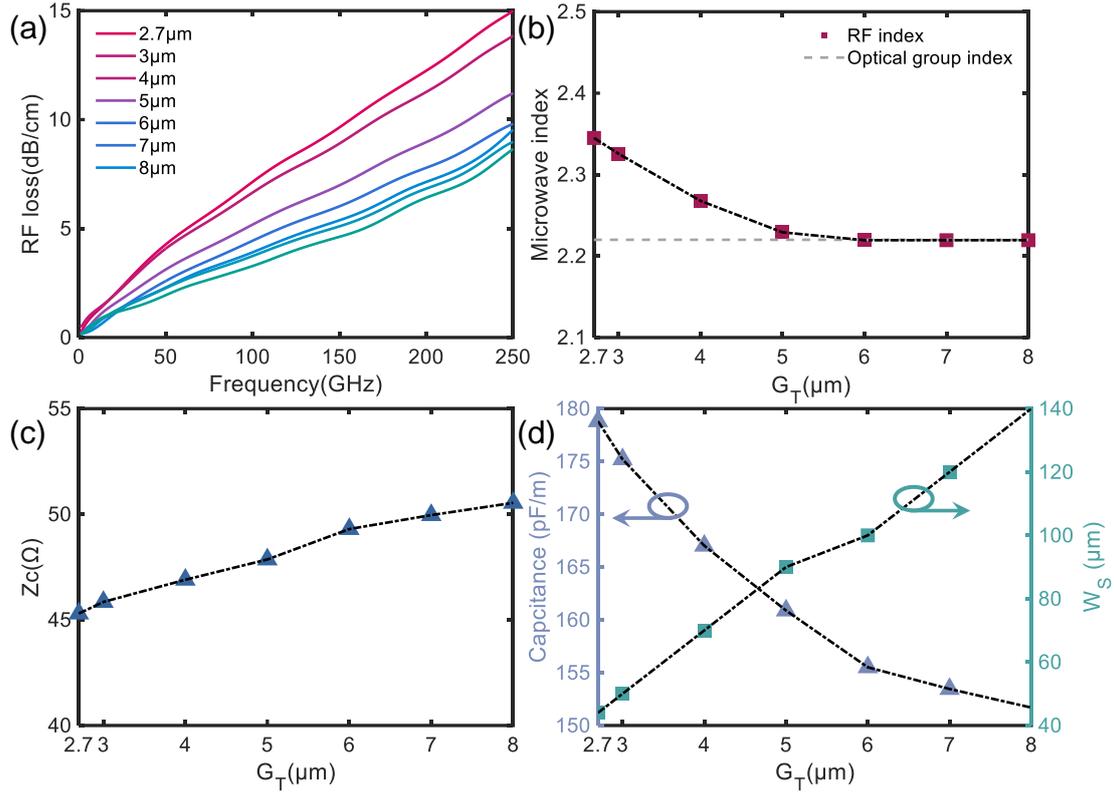

Figure 3. The influences of electrode gap $G_T$ on (a) microwave loss, (b) microwave index, (c) characteristic impedance, (d) capacitance and signal electrode width.

To overcome the hurdle of the large capacitance associated with a narrow T-rail gap, low dielectric constant ($\varepsilon$ = 2.56) BCB underfill is formed beneath the main CPW electrodes, as shown in Figure 1(b). The introduction of low-k BCB underfill reduces the effective dielectric constant, which corresponds to a smaller distributed capacitance and allows for a wider signal electrode, thus leading to reduced electrode resistance and microwave loss. Meanwhile, it also helps realize velocity matching, as the effective microwave refractive index can be tuned by adjusting the thickness of the BCB underfill.

Figure 4(a) shows the simulated microwave loss of CL-TWEs on quartz substrate with different BCB underfill thickness, assuming a fixed electrode gap of 2.7 μm. The structural parameters of CL-TWEs with different BCB underfill thickness are optimized to secure both velocity and impedance matching while minimizing the microwave loss. The dielectric loss is taken into account by using the loss tangent data summarized in Table I. As revealed by Figure 4(a), the introduction of BCB layer results in remarkable reduction of microwave loss. Figure 4(b) plots the calculated microwave effective refractive indices of CL-TWEs on quartz substrate with and without BCB underfill. The BCB underfill helps reduce the microwave refractive index, thus facilitating velocity matching. Ultimately, taking the fabrication feasibility into account, a 2-μm-thick BCB layer together with an 85-μm-wide signal line is adopted for reduced microwave loss. In passing, it is noted that the low-k underfill can also be employed in TFLN modulators formed on silicon substrate for reducing RF loss and accomplishing velocity matching, thus avoiding the complicated substrate removal process to form a suspended structure [16,17].

Table I. Dielectric Loss Parameters Used in Simulation

| Materials | LN | PECVD-Silica | Quartz | BCB |
| --- | --- | --- | --- | --- |
| Loss tangent | 0.008 | 0.006 | 0.0004 | 0.002 |

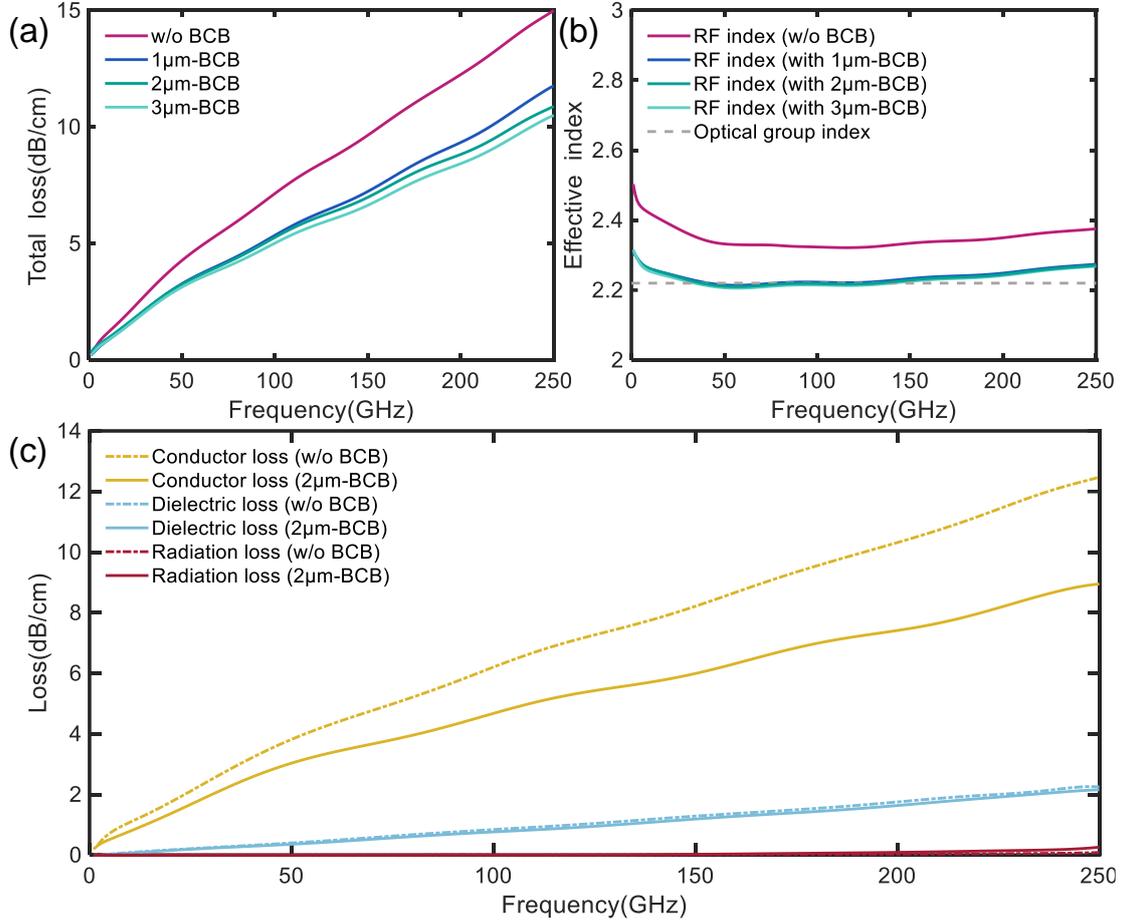

Figure 4. (a) Microwave losses and (b) effective microwave indices of different CL-TWEs. (c) The simulated loss of CL-TWEs on quartz substrate with and without 2-μm-thick BCB underfill.

The microwave loss of the CL-TWEs consists of contributions with different frequency dependence. Previous investigations have shown that the conductor loss due to the main CPW electrodes is proportional to the square root of frequency [18], whereas that due to the T-rails is proportional to the square of frequency [19]. The dielectric loss is linearly related to frequency, while the radiation loss varies cubically with frequency [15,20]. As a result, the frequency dependance of the total loss can be modeled as

$$\alpha_{total} = \alpha_{cpw}\sqrt{f} + \alpha_d f + \alpha_T f^2 + \alpha_R f^3 + \alpha_{dc} \tag{4}$$

where $\alpha_{dc}$ represents a frequency independent loss. Figure 4(c) shows a breakup of the simulated microwave losses for CL-TWEs with and without 2-μm-thick BCB underfill. It is found that the conductor loss, due to both the main CPW electrodes and the T-rails, is the dominant contribution, whereas the radiation loss has a negligible impact. A 2-μm-thick BCB underfill helps reduce the conductor loss effectively, thus leading to significant total microwave loss reduction.

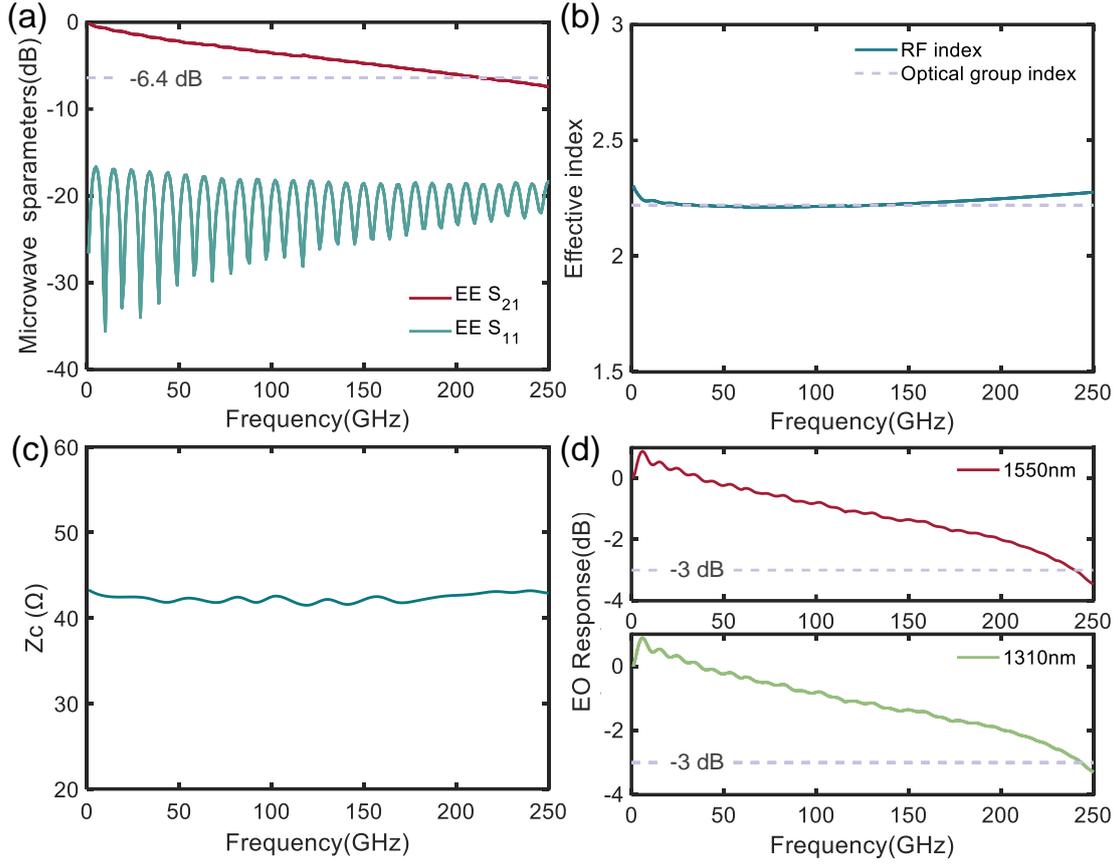

Figure 5. (a) Simulated microwave transmission and reflection. (b) Extracted microwave effective index and optical group index. (c) Extracted characteristic impedance. (d) Calculated electro-optic response of the TFLN modulator in the C- and O-band.

The simulated microwave S-parameters for a TFLN modulator incorporating 7-mm-long CL-TWEs with 2.7-μm-wide T-rail gap and 2-μm-thick BCB underfill are illustrated in Figure 5(a), indicating a 6.4-dB electrical bandwidth about 220 GHz and good impedance matching with $S_{11}$ below −17 dB. Figure 5(b) shows the extracted microwave effective index, which is close to the optical group index ($n_g$ = 2.221@1550nm and 2.225@1310nm). Based on these results, the electro-optic modulation response can be calculated according to the EO frequency response model [21], as shown in Figure 5(d). The 3-dB EO modulation bandwidth is estimated to be about 240 GHz (244 GHz) in the C-band (O-band).

## 3. Device Fabrication and Characterization

### 3.1 Device Fabrication

The TFLN wafer is first patterned by electron beam lithography (EBL), and 1-μm-wide and 200-nm-thick LN waveguides are formed by reactive-ion etching (RIE) with argon. A 100-nm-thick $SiO_2$ layer is then deposited over the TFLN wafer by plasma-enhanced chemical vapor deposition (PECVD). To ensure high alignment accuracy, T-rail patterns with 2.7 μm narrow gap are defined by EBL, and the 200-nm-thick Cr/Au T-rail electrodes are formed via magnetron sputtering and bilayer resist lift-off. Subsequently, a 2-μm-thick photosensitive BCB layer is spin-coated onto the wafer and patterned by UV lithography to expose the T-rail electrodes for contact with the main electrodes. After curing the BCB film in vacuum, main CPW electrodes are formed by magnetron sputtering and thickened to 1.7 μm by electroplating for reduced microwave loss. An on-chip NiCr resistor is formed at the end of the CL-

TWEs to terminate the microwave modulation signal. Figure 6(a) shows the optical microscope image of the fabricated modulator with a 7 mm modulation length, while Figures 6(b) and 6(c) show a detail of the modulation area and the on-chip terminal resistor.

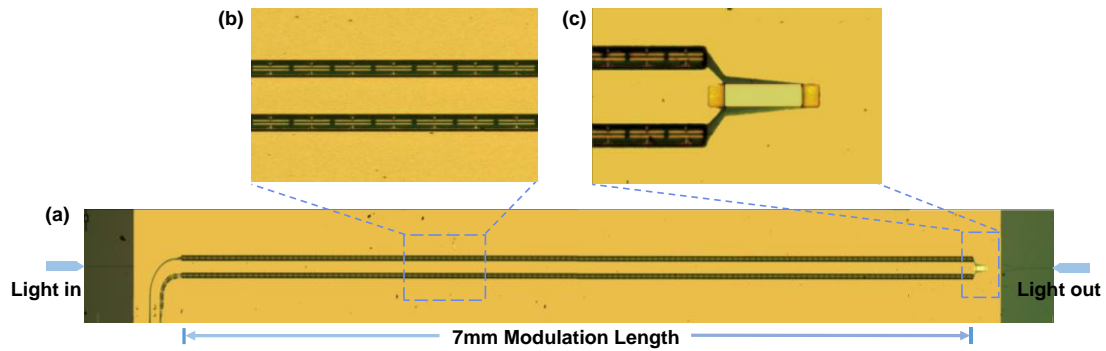

Figure 6. Optical microscope images of the (a) fabricated modulator with 7 mm modulation length; (b) modulation area and (c) the on-chip terminal resistor.

*3.2 Microwave S-Parameters Characterization*

Figure 7(a) shows the S-parameters of a 7-mm-long CL-TWEs measured by a 110 GHz vector network analyzer (VNA, Keysight N5291A) and a pair of 110 GHz GSG microwave probes. The RF loss is only about 3.4 dB up to 110 GHz, revealing the low microwave loss property of the CL-TWEs with low-k underfill. The extracted microwave index is plotted in Figure 7(b), and perfect velocity matching to the optical group index is confirmed. The microwave index is 2.22 at 110 GHz, in excellent agreement with our simulation shown in Figure 5(b). Figure 7(d) compares the microwave losses of the fabricated devices, revealing significant reduction in microwave loss for CL-TWEs with low-k underfill. To the best of our knowledge, our structure exhibits the lowest microwave loss among TFLN modulators reported so far.

Based on the measured microwave S-parameters and the analytic model given in Equation (4), the different loss contributions are extracted and shown in Figure 7(e). The conductor loss dominates the overall losses. At frequencies below 100 GHz, the primary contribution comes from the CPW loss, while at frequencies above 150 GHz, the T-rail loss becomes dominant.

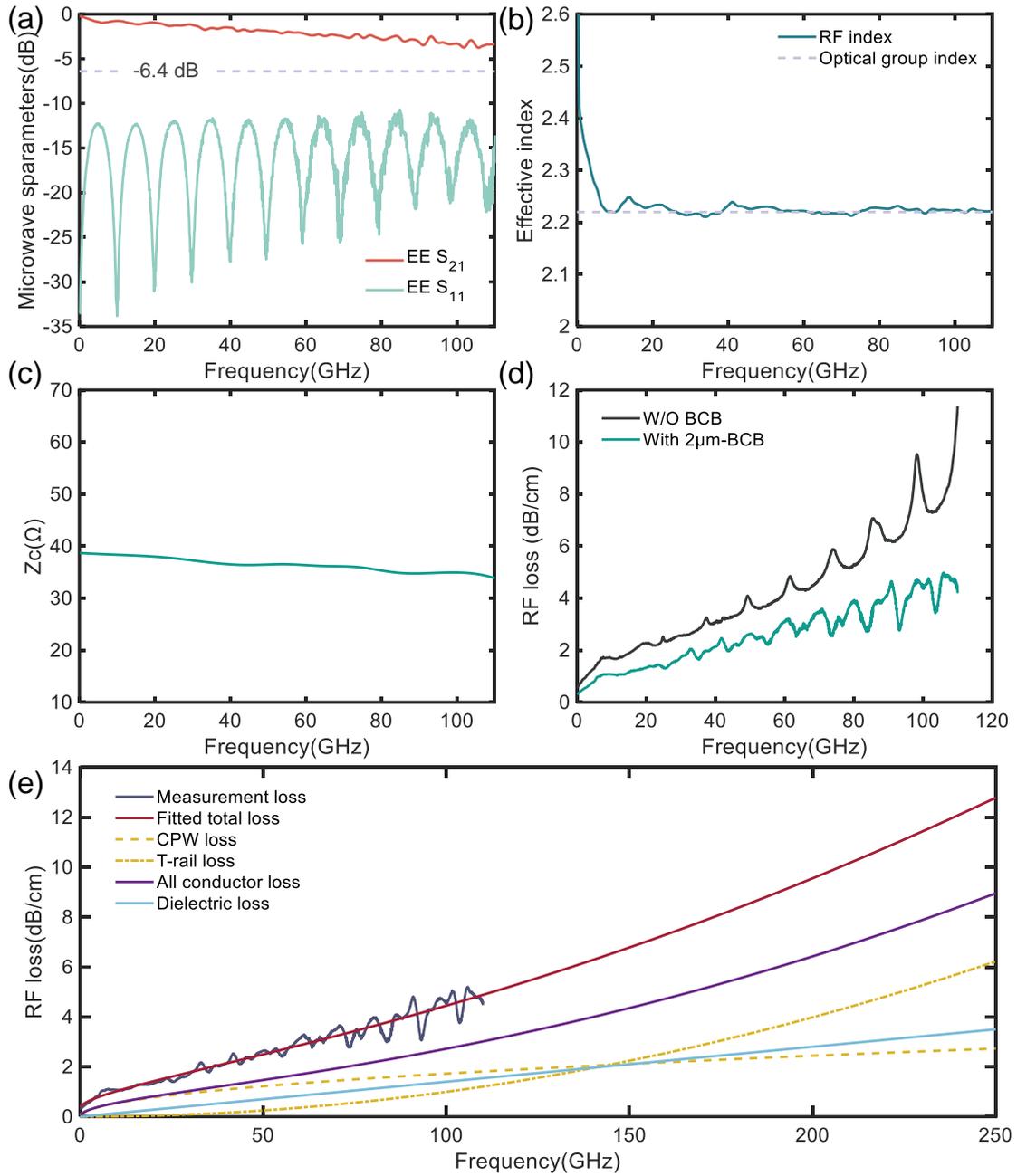

Figure 7. (a) Measured S-parameters of a modulator with 7 mm modulation length. Extracted (b) microwave refractive index and (c) characteristic impedance. (d) Microwave losses with and without low-k underfill. (e) Fitted microwave loss, CPW conductor loss, T-rail conductor loss and dielectric loss as a function of frequency.

*3.3 Electro-optic Response Characterization*

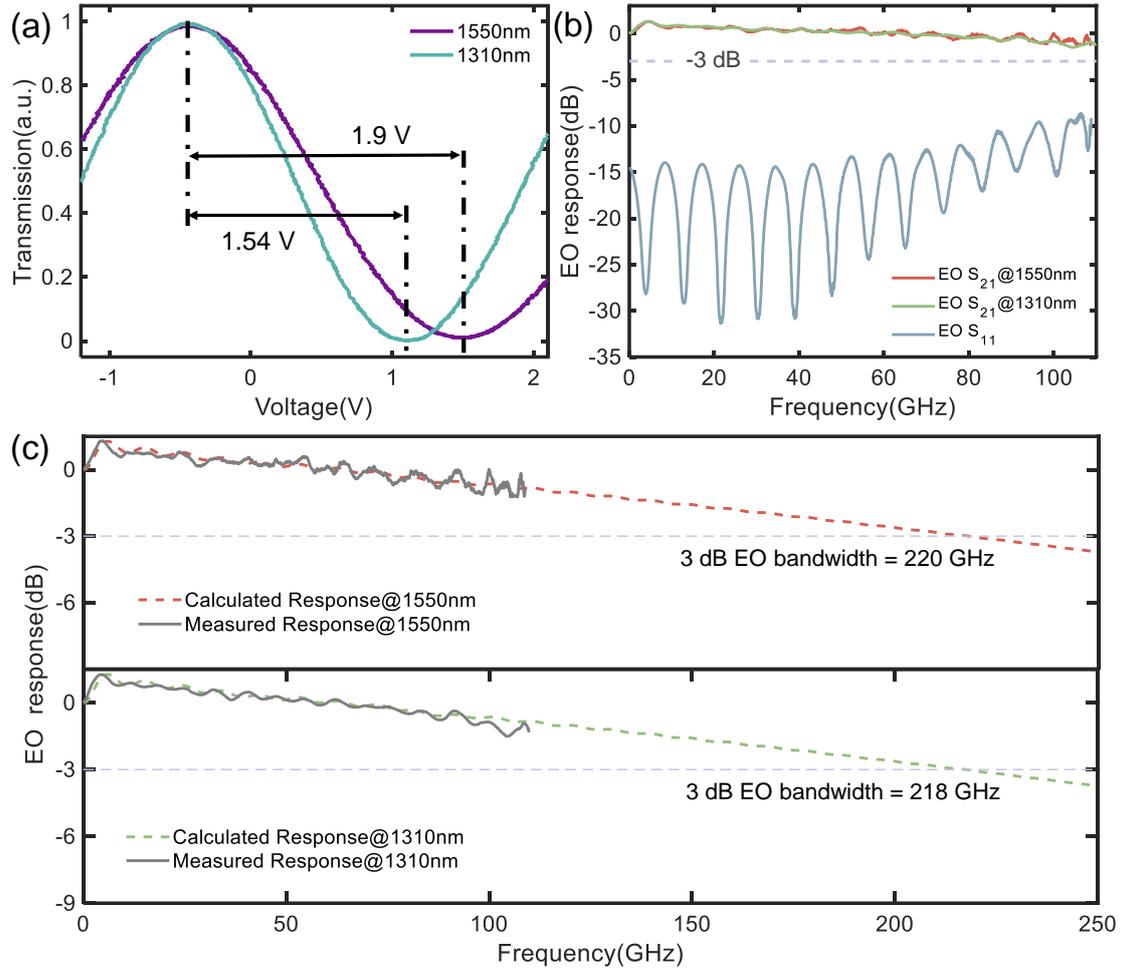

Figure 8. (a) Extinction behavior of a 7-mm-long modulator. (b) Measured electro-optic response and electrical reflection up to 110 GHz. (c) Measured and predicted electro-optic responses up to 250 GHz at 1550 nm and 1310 nm.

Figure 8(a) plots the modulation curves recorded at 1 MHz, indicating a $V_\pi$ of 1.9 V at the C band, corresponding to a low $V_\pi L$ of 1.33 V·cm, in fair agreement with our theoretical estimation. The EO response of the modulator is characterized by a 110 GHz lightwave component analyzer (LCA, Keysight N4372E), as illustrated in Figure 8(b). An on-chip termination resistor of 35 Ω, which is slightly lower than the characteristic impedance of the CL-TWEs, is adopted to eliminate the steep drop at low frequencies [22]. As depicted in Figure 8(b), the EO frequency response exhibits a 3-dB EO bandwidth far beyond 110 GHz, while the electrical reflection $S_{11}$ remains below −10 dB. Based on the extracted RF loss, effective microwave index and characteristic impedance, the EO response of our modulator is extrapolated beyond 110 GHz by using the analytical model of traveling-wave MZMs [21]. As illustrated by Figure 8(c), the simulation and experimental results match perfectly up to 110 GHz, and the extrapolated 3 dB bandwidth of our modulator is 220 GHz with a roll-off less than 0.77 dB up to 110 GHz.

We have also tested the performance of the same modulator at the O-band. According to Figure 8(a), the modulator exhibits a halfwave voltage of 1.54 V at 1310 nm, corresponding to a $V_\pi L$ of 1.08 V·cm. Meanwhile, as revealed by Figures 8(b) and 8(c), the roll-off in the EO response at 110 GHz is only 0.83 dB for the O band, with an extrapolated 3 dB bandwidth of 218 GHz. The modulator exhibits the same outstanding performance in both C- and O- bands, thanks to the nearly identical optical group velocities for the two bands ($n_g$ = 2.221@1550 nm and 2.225@1310 nm).

## 4. High-speed Data Transmission Demonstration

To further verify the high-speed data transmission capacity of our wide-bandwidth modulator, a series of eye diagram tests under non-return-to-zero (NRZ) and pulse amplitude modulation (PAM) are carried out in the C- and O-bands, with the experimental setup illustrated in Figure 9(a). At the transmitter end, a pseudorandom bit sequence (PRBS) is generated by a 256 GSa/s AWG (Keysight M8199A) and then mapped to NRZ/PAM symbols. After resampling the symbols to the digital-to-analog converter (DAC) sampling rate, the pulse signals are shaped with a raised cosine filter. Next, a pre-emphasis filter pre-compensates for the low pass filtering of the DAC and 67 GHz electrical amplifier (EA, GT-LNA-67G). It should be noted that we did not employ any nonlinearity pre-compensation in our experiment. The amplified RF signal and the bias voltage are combined by a 65 GHz Bias-T (SHF BT652-B) and then fed to the modulator via a 67 GHz GSG probe. An erbium-doped fiber amplifier (EDFA) or a praseodymium-doped fiber amplifier (PDFA) is used to boost the modulated optical signal. At the receiver end, the transmitted optical signal is captured by a digital sampling oscilloscope (DSO) equipped with a 65 GHz optical sampling module (Keysight N1030A) and then processed using simple low-pass filtering and feed-forward equalization (FFE).

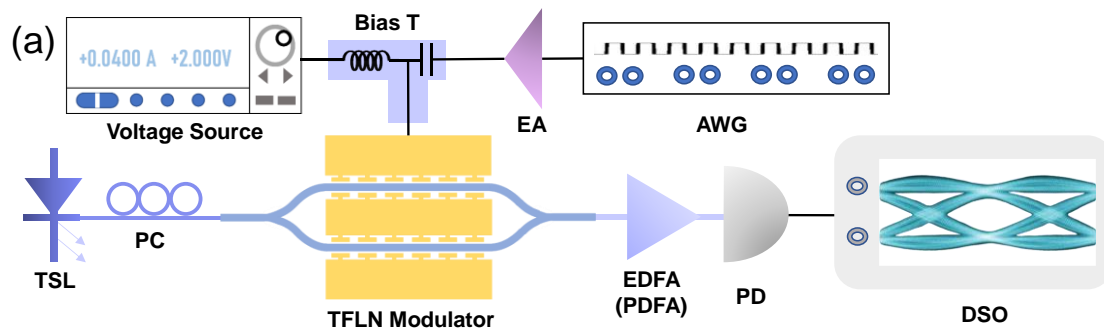

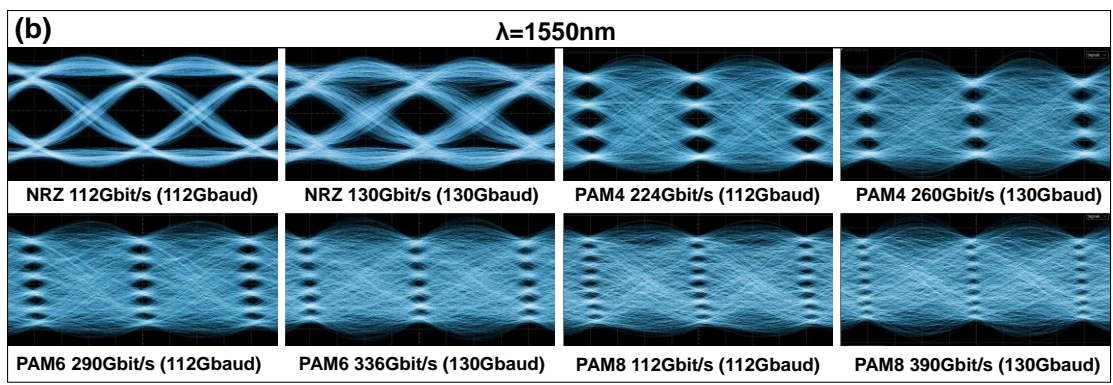

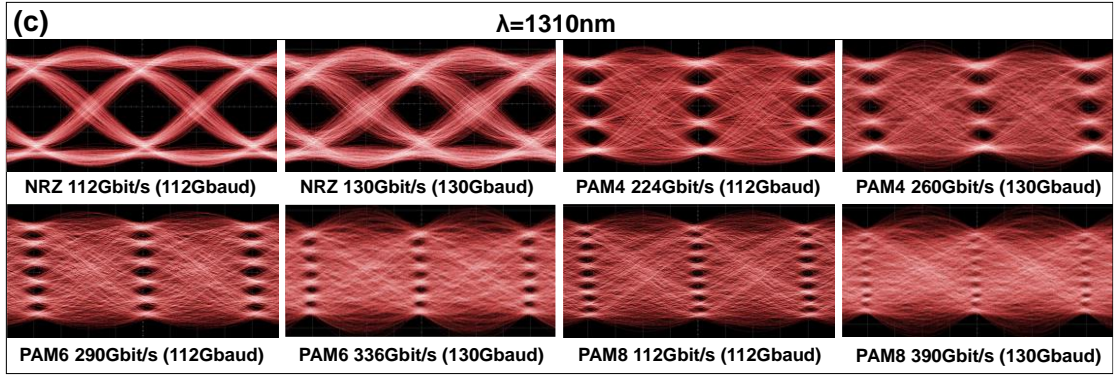

Figure 9. (a) Schematic of the setup for high-speed eye diagram measurements of the TFLN modulator. Measured eye diagrams with different symbol rates and modulation formats at (b) 1550 nm and (c) 1310 nm. TLS: tunable laser source, PC: polarization controller, AWG: arbitrary waveform generator, EA: Electrical amplifier, EDFA: erbium-doped fiber amplifier, PDFA: praseodymium-doped fiber amplifier, PD: photodetector, DSO: digital sampling oscilloscope.

Figures 9(c) and 9(d) depict the recorded optical eye diagrams for NRZ, PAM4, PAM6 and PAM8 modulation format with symbol rates of 112 Gbaud and 130 Gbaud at 1550 nm and 1310 nm. Clear eye opening has been demonstrated at different symbol rates for both C- and O bands. Currently, the data rate is mainly limited by the analog bandwidth of the RF components used in the transmission system, including AWG, EA, Bias-T, microwave probes and photodetector (PD). We firmly believe that our high-performance modulator would support a higher single-lane data rate by employing an AWG with higher baud rate ability in our future demonstration.

The energy consumption per bit by the modulator can be estimated as $W_e = V_{rms}^2/(B \cdot R)$ [3], where $V_{rms}$ is the root-mean-square drive voltage, B is the bit-rate and R is the driver impedance. In our experiment, an electrical $V_{rms}$ of 245.6 mV (97.6 mV) is used for the 390 Gbit/s data modulation in the C-band (O-band), corresponding to an electrical energy consumption of 4.42 fJ/bit (0.69 fJ/bit) in the C-band (O-band). To the best of our knowledge, this is the lowest power consumption reported in IM-DD transmission experiments. It is noted that the energy consumption also includes the power consumption of the microwave probe and the RF cable.

Table II. Performance comparison of TLFN modulators with IM-DD demonstration in the C- and O-bands

| References | Optical Band | $V_\pi$ [V] | $V_\pi L$ [V·cm] | 3-dB BW [GHz] | Data Rates [Gbit/s] | Energy Consumption [fJ/bit] | Electrode Structure |
|---|---|---|---|---|---|---|---|
| 2018 [3] | C | 1.4 | 2.8 | > 45 | 210 (PAM8) | 14 | CPW |
| 2019 [23] | C | 5.1 | 2.5 | 70 | 112 (PAM4) | 170 | CPW |
| 2020 [24] | C | 3 | 2.7 | > 67 | 128 (PAM4) | / | CPW |
| 2022 [17] | C | 2.2 | 2.2 | > 67 | 112 (PAM4) | / | CL-TWEs |
| 2022 [5] | C | 4.7 | 2.37 | 110 | 250 (PAM6) | / | CPW |
| 2022 [7] | C | 4.4 | 2.2 | 84 | 240 (PAM8) | 147 | CPW |
| 2023 [25] | C | 14.8 | 2.96 | 50 | 70 (OOK) | / | CPW |
| 2023 [13] | C | 3.52 | 1.41 | > 67 | 64 (OOK) | / | CPW+High-k |
| 2024 [26] | C | 2.18 | 2.18 | > 67 | 112 (PAM4) | / | CL-TWEs |
| 2024 [27] | O | 2.04 | 1.02 | 108 | 224 (PAM4) | / | CPW+ITO |
| **This work** | **C** | **1.9** | **1.33** | **> 110 (220[a])** | **390 (PAM8)** | **4.42** | **CL-TWEs+BCB** |
| **This work** | **O** | **1.54** | **1.08** | **> 110 (218[a])** | **390 (PAM8)** | **0.69** | **CL-TWEs+BCB** |

[a]: Extrapolated 3-dB EO bandwidth.

Table II summarizes the performances of TFLN MZMs reported in IM-DD transmission. Our modulator exhibits the same outstanding overall performance of low $V_\pi$, wide EO bandwidth and ultra-high modulation efficiency in both C- and O-bands. And ultra-high-speed data transmission for IM-DD system with lowest energy consumption in dual bands has also been demonstrated.

Compared to TFLN modulators employing high-k cladding [13] or transparent conductive oxide (TCO) electrodes [27] for enhanced modulation efficiency, the proposed modulator incorporating low-k BCB underfill is simple to implement and suitable for low-cost massive production. The high modulation efficiency allows for decreased device length, making it attractive for compact package, such as quad small form factor pluggable double density (QSFP-DD). The over 200 GHz potential bandwidth can support future data transmission with ultra-high baud rates beyond 200 Gbaud. Dual band operation capability of the modulator enables its application in both the long-haul fiber communication in the C-band or short-reach interconnections in the O-band with high wavelength multiplexing capability.

5. Conclusion

In this work, we have experimentally demonstrated a high-performance dual-band TFLN MZM. A low-k BCB underfill is introduced to markedly improve the electro-optic bandwidth while maintaining high modulation efficiency with a simple electrode configuration. Ultra-low microwave loss and perfect velocity matching are secured for traveling-wave electrode with narrow electrode gap by forming BCB underfill, thereby improving the modulation efficiency and high-frequency EO performance at the same time. The fabricated 7-mm-long modulator exhibits a low $V_\pi L$ of 1.33 V·cm (1.08 V·cm) and ultra-flat EO response with a roll-off less than 0.77 dB (0.83 dB) up to 110 GHz at the C-band (O-band), along with extrapolated bandwidth about 220 GHz. To the best of our knowledge, this is the first demonstration of high-speed data transmission using the same modulator in both bands. Up to 390 Gbit/s data rate with 130 Gbaud PAM8 modulation with extremely low energy consumption as low as sub-pJ/bit level is demonstrated for the next-generation ultrahigh-speed and low-power IM-DD communication systems [28]. The novel low-k design pushes the overall performance of TFLN modulators beyond previous limits, and our device is anticipated to hold significant potential for future terabit-per-second optical communication applications featuring energy-efficient, low cost, and multi-wavelength support.

**Acknowledgement**

This work was supported in part by National Key R&D Program of China (2022YFB2803002); National Natural Science Foundation of China (62235005, 62127814, 62225405, and 62475130); and Collaborative Innovation Centre of Solid-State Lighting and Energy-Saving Electronics.